# Topologically-protected superluminal pair annihilation in photonic time crystals

Liang Zhang[1], Chenhao Pan[1], Jinze He[1], Danni Chen[1], Zirui Zhao[1], Qingqing Cheng[2,*], Yiming Pan[1,*]

[1]State Key Laboratory of Quantum Functional Materials, School of Physical Science and Technology and Center for Transformative Science, ShanghaiTech University, Shanghai 200031, China

[2]Shanghai Key Laboratory of Modern Optical System, School of Optical-Electrical and Computer Engineering, University of Shanghai for Science and Technology, Shanghai 200093, China.

*yiming.pan@shanghaitech.edu.cn

**Abstract**

Photonic time crystals (PTCs) - dielectric media whose permittivity is periodically modulated in time - map to a Dirac equation with an imaginary mass, opening a momentum gap (k-gap) where modes grow or decay exponentially. Here, we introduce a sequence of temporal Jackiw-Rebbi kinks that act as a programmable flip of the Dirac mass, exchanging the amplifying and decaying in-gap modes. By launching two seeded pulses with a controlled relative phase, we demonstrate topological pair annihilation in spacetime domain, the phase-selective cancellation of counter-propagating, k-gap-amplified modes. The resulting spatiotemporal cascade appears superluminal, yet causality is preserved because the cascaded pattern carries no net energy flux. To facilitate implementation, we construct a minimal time-varying non-Hermitian lattice model and reproduce the phase-selective pair annihilation behavior, establishing a direct continuum-lattice correspondence. Our results identify topological kinks as temporal gating to manipulate the growth and wave propagation of time-varying media.

Photonic time crystals (PTCs) are optical media with refractive index modulated periodically in time. A defining feature of PTCs is the emergence of momentum (k) gaps [1,2], where the Floquet quasifrequency becomes complex (having both real and imaginary parts). Modes inside these k-gaps either grow or decay exponentially by exchanging energy with the modulation. The growing branch has been proposed for lasing and amplification [3,4], and because the medium is spatially uniform, these modes appear in pairs (as the forward propagating and time-reflected modes). For wave manipulation, a central challenge is therefore to regulate the growth of k-gap modes in a controlled manner.

Early approaches relied on loss, which suppresses amplification but also removes interesting dynamics. Nonlinear effects provide a gentler route: Pan et al. showed that adding Kerr nonlinearity prevents runaway growth and yields exotic states such as superluminal k-gap solitons and event solitons [5,6]. A complementary idea is to actively excite the decaying modes that coexist with the growing modes inside the k-gap, so that k-gap amplification can be actively balanced by these decaying modes. For instance, in 2018 Lustig et al. demonstrated that a temporal kink can transiently convert an amplifying in-gap mode into a decaying one without altering its momentum [7], forming temporal topological modes and providing a powerful knob. Such temporal topological modes can be understood through an extended Jackiw-Rebbi framework [8–10].

These phenomena can be organized within a massive Dirac equation. Unlike a conventional energy gap with a real Dirac mass, temporal modulation in PTCs produces an imaginary mass that opens momentum gaps. Prior works [7,11–13] described temporal topological modes using a Dirac equation with kink-profiled imaginary mass in time. In the PTC setting, this corresponds to converting a forward-propagating (+k) growing mode into a forward decaying mode, and likewise for the backward-propagating (-k) partner. Because the modulation is temporal but spatially uniform, momentum is conserved. This explains why the temporal Jackiw–Rebbi solutions appear in pairs for PTCs, and the resulting pairwise conversion naturally admits an operational interpretation as pair annihilation [5,14,15].

Building on this understanding, we consider two independently seeded pulses. Each seed, when amplified inside the k-gap, generates forward- and backward-propagating partners. Since temporal kinks act uniformly across momentum, the backward mode of the front pulse inevitably overlaps with the forward mode of the back pulse. This overlap enables pair annihilation between different seeds. Crucially, such annihilation is controlled by the relative phase of the seeds, which decides whether the overlapping modes annihilate or reinforce. To examine this phase-selective pair annihilation, we design a sequence of temporal kinks that repeatedly flip the imaginary mass and stabilizes otherwise unbounded k-gap growth. With single-pulse seeding, this kink protocol generates the characteristic superluminal triangular spatiotemporal patterns; with two seeds of controlled phase and locations (as shown in Fig. 1) yields local pair generation and annihilation within their overlapping region. This annihilation is strictly phase-dependent, arising from spatiotemporal interference. Since direct testing of PTCs requires ultrafast modulation in homogeneous media that is challenging, we also construct a simple bipartite lattice model with time-varying staggered gain/loss profile. This non-Hermitian lattice realizes temporal kinks through the gain/loss modulation and reproduces the k-gap-amplified spatiotemporal patterns and phase-selective pair annihilation. In both PTC and discrete-lattice settings the Dirac equation with a time-modulated imaginary mass term provides a unified description, establishing topological kinks as temporal gates for programmable wave control in time-varying systems.

**Equivalence of k-gap in PTC with the imaginary mass term in Dirac equation.**
We begin with the continuum description of a photonic time crystal (PTC), a spatially homogeneous medium whose permittivity is modulated periodically in time. Starting from Maxwell's equations and applying the slowly varying envelope approximation for the forward- and backward-propagating components, the dynamics reduce to a two-component Dirac equation (see Methods for the derivation):

$$i\partial_t \psi = -ic\sigma_z \partial_x \psi - i\kappa c^2 \sigma_y \psi, \qquad (1)$$

where $\sigma_{z,y}$ are Pauli matrices, $c = c_0/\sqrt{\epsilon_r} = c_0/n_0$ is the light speed in the medium, and $\kappa = \delta\Omega/8c^2$ is a temporal modulation coefficient defined by the modulation amplitude $\delta$ and frequency $\Omega$. Unlike the standard massive Dirac Hamiltonian, the mass term in the PTC appears as an imaginary contribution $-i\kappa$, which fundamentally modifies the band structure. Seeking plane-wave solutions $\psi = \chi e^{ikx-i\omega t}$ yields the eigenvalue equation $\omega\chi = (ck\sigma_z - i\kappa c^2 \sigma_y)\chi$. This leads directly to the hyperbolic dispersion $\omega^2 = c^2 k^2 - (\kappa c^2)^2$. Equivalently, $k_\pm = \pm\sqrt{(\omega/c)^2 + (\kappa c)^2}$, which shows that a momentum gap ($k$-gap) opens for $k < |\kappa c|$, with width $2\kappa c = \delta\Omega/4c$. Inside the $k$-gap, $\omega$ is purely imaginary and the modes are unstable in time, growing or decaying exponentially by exchanging energy with the modulation. The growth or decay rate $Im\,\omega$ increases with the modulation coefficient $\kappa$ (aka, with the gap size). To this end, for bandgap engineering a real Dirac mass term opens an energy ($\omega$) gap, whereas an imaginary mass opens a momentum ($k$) gap.

In the 1D Dirac model with real mass, the energy-gapped dispersion reads $E_\pm = \pm\sqrt{(vk)^2 + m_D^2}$. When the mass $m_D(x)$ varies spatially and reverses sign at $x_0$ (a domain wall, DW), the gap closes at $k = 0$ as $m_D \to 0$ and reopens with inverted band order. This band inversion supports a topological zero mode - the Jackiw–Rebbi solution [8]. In the adiabatic limit, the bound state takes the form $\psi_0 \propto \exp\left[-\frac{m_0|x-x_0|}{v}\right]$, pinned at $E = 0$ and localized near the wall [Fig. 1a]. Across the interface, the bulk Zak phase differs by $\pi$ (a polarization shift of $e/2$), giving this mid-gap state by bulk-edge correspondence.

For PTCs, we invoke the space-time analogy by replacing x with t. The Dirac mass becomes imaginary and time-dependent, $-i\kappa(t)$. A temporal kink is created by flipping its sign at $t = T_0$ (forming a temporal domain wall), a smooth profile is $\kappa(t) = \kappa_0 \tanh(t - T_0)$, which gradually inverts the sign [Fig. 1b]. In the sharp limit, $tanh \rightarrow sgn$, enforcing discontinuity and time boundary at $t = T_0$, it yields a Jackiw-Rebbi-like temporal boundary state $\psi \propto e^{-\Lambda|t-T_0|}$. The localization rate $\Lambda$ is set by the local k-gap parameter, equivalently $\Lambda = \text{Im}\,\omega(k)$, and in PTCs scales as $\Lambda \propto \kappa_0$, just as the spatial localization is controlled by the real mass $m_D$. The key difference from the spatial DW is that the temporal topological state is transient: it nucleates at the temporal kink and subsequently grows and decays in time, reflecting the unstable nature of k-gap modes in PTCs.

Inside the $k$-gap, the dynamics is governed by two solutions: a growing mode $e^{+|\text{Im}\,\omega|t}$ and a decaying mode $e^{-|\text{Im}\,\omega|t}$. A temporal kink flips the sign of $\kappa$, exchanging their stability and converting a unstable growing solution into a decaying one. This yields a time-localized "grow-then-damp" response centered at the kink; far from the interface, the amplified modes is reselected and exponential growth resumes [7]. For a single forward-propagating input with wavevector $k_{\Omega/2}$ lying within the k-gap, momentum conservation enforces creation of a counter-propagating partner at $-k_{\Omega/2}$. This constitutes pair generation with symmetric branches at $\pm k_{\Omega/2}$ [Fig. 1c]. The temporal Jackiw–Rebbi solution likewise appears in pairs: a kink launches two time-localized, spatially finite wavepackets carrying opposite momenta. [Fig. 1d]. With multiple kinks, each newly generated packet seeds further splitting at the next kink, yielding a cascade of spatiotemporal patterns.

While pair generation is a generic feature of k-gap physics, temporal kinks provide active control by converting seeds into their decaying counterparts and thereby suppressing the infinite amplification. This renders both pair generation and annihilation experimentally accessible and raises the central question: can pair annihilation occur not only within a single generated pair, but also between modes seeded from different sources? In particular, how does this process depend on the relative phase between the seeds?

**Pair annihilation in continuous PTC model.** To model pair annihilation, we consider a continuum PTC whose forward and backward wave envelopes obey Eq. (1): a Dirac equation with a time-dependent imaginary Dirac mass $-i\kappa(t)$. We impose a train of temporal kinks on the mass profile:

$$\kappa(t) = \kappa_0 \prod \tanh[10(t - T_0 - n * T_{rep})], \qquad (2)$$

where $T_0$ is the time of the first kink and $T_r$ is the repetition period. This construction realizes a sequence of controlled mass sign reversals that swap the growing and decaying solutions at prescribed instants. For simulations, we set the base modulation frequency to $\Omega = 18\pi$ and use a modulation depth $\delta = 0.3$, which gives $\kappa_0 = \delta\Omega/8c^2 \approx 0.675\pi$ (with $c = 1$). These parameters expand the k-gap, making the growth and decay rates of k-gap modes resolvable within short temporal windows.

To probe pair annihilation between two different seeds, we launch two identical Gaussian wavepackets from $x_0 = \pm 4.5$ with initial phases $\phi_1, \phi_2$ (phase difference $\Delta\phi = \phi_1 - \phi_2$) and spectra confined to the $k$-gap. Each seed crossing a temporal kink generates an amplified, counter-propagating partner with opposite momentum. The initial positions are chosen so that the $+k_{\Omega/2}$ component from one seed spatiotemporally overlaps the $-k_{\Omega/2}$ component from the other. This configuration defines a clean triangular overlap region (TOR) that isolates phase-dependent interference and allows direct observation of pair annihilation.

Each temporal kink at $t = T_0 + nT_r$ creates topological wavepackets, initiating a kink-to-kink scattering cascade. The wavefront of forming patterns exhibit an apparently superluminal advance ($v \approx 1.5c$ in Fig. 2a). This effect does not violate causality: in-gap modes are temporally unstable and carry no net energy flux [16]. Such superluminal fronts have been reported previously, for example in studies of superluminal k-gap solitons [5]. Thus, the observed triangular cascades reflect how temporal kinks regulate pair generation and annihilation seeding from a single pulse. To study pattern interference starting from two seed pulses, we zoom into the overlap region marked by a red dashed triangle. Within this TOR, the counter-propagating partners from the two seeds overlap and interfere. For a phase difference $\Delta\phi = \pi/2$, the TOR intensity follows the linear superposition baseline [Fig. 2a]. In contrast, for $\Delta\phi = \pi$ in Fig. 2b, the pattern is strongly depleted: the overlapping $\pm k$ modes undergo pair annihilation and the intensity nearly vanishes, while regions outside the TOR continue their superluminal cascade.

The global power $P(t) = \int |E(x,t)|^2 \, dx$ captures the cumulative effect of kink-induced dynamics [Fig. 2c]. The orange trace at the top marks the temporal kink train, and the circles indicate step-like increments triggered by each kink. For $\Delta\phi = \pi/2$ (brown), the energy follows the red dashed guide, an overall exponential envelope - with a reduced growth rate compared to a purely periodic drive (dashed gray line), providing direct evidence of gain gating by temporal kinks. For $\Delta\phi = \pi$ (blue), the post-overlap growth is markedly slower than in the $\Delta\phi = \pi/2$ case. This pronounced suppression demonstrates phase-selective pair annihilation within the TOR.

To quantify phase control, we define the pair-annihilation efficiency inside the red triangular overlap region:

$$\eta_{PA}(\Delta\phi) = 1 - \frac{\iint_C I(x,t,\Delta\phi) \, dxdt}{\iint_C I_L(x,t) + I_R(x,t) \, dxdt} \quad (3)$$

where $I(x,t;\Delta\phi)$ is the measured intensity with both seed pulses present, and $I_L, I_R$ are reference for each seed alone. By definition, $\eta_{PA} > 0$ signals pair annihilation, $\eta_{PA} = 0$ corresponds to linear addition, and $\eta_{PA} < 0$ indicates excess generation. The phase scan in Fig. 2d (blue dots) shows $\eta_{PA}$ peaking near $\Delta\phi = \pi$ with near-complete annihilation, and dipping near $\Delta\phi \approx 0$ for constructive outcomes, confirming strict phase selectivity. This arises because the two seeds retain their initial phase while evolving in the k-gap, which determines the spatiotemporal patterns in the TOR.

**Realization in a lattice model.** Direct test of the continuous PTC model requires ultrafast temporal modulation and is experimentally challenging. To gain feasibility, we consider discrete implementations where in k-gap physics is already realized, such as photonic circuits [17,18], acoustics [19,20], and synthetic-dimension platforms [11,12]. Over a finite Bloch-momentum window, these systems are captured by the same Dirac description used above. Accordingly, we construct a minimal bipartite chain (Fig. 3a) that retains three ingredients: conserved Bloch momentum, non-Hermitian exchange via balanced gain/loss, and a temporally kinked control parameter. The real-space Hamiltonian reads

$$H_0 = \sum_{j=1}^{N} (i\gamma(t) c_{jG}^\dagger c_{jG} - i\gamma(t) c_{jL}^\dagger c_{jL}) + \sum_{j=1}^{N-1} \tau(c_{jG}^\dagger c_{jL} + c_{jL}^\dagger c_{j+1G}) + h.c., \quad (4)$$

where $\tau$ is the intersite coupling and $\gamma(t)$ is the effective time-dependent balanced gain/loss, serving as the analogue of the temporal kink sequence that implements the mass inversion in the continuous model. In momentum space, the Hamiltonian is $H_0 =$ $\sum_k H_k C_k^\dagger C_k$, where $C_k = (c_{kG}, c_{kL})^T$, with $H_k = 2\tau \cos k\, \sigma_x - i\gamma(t)\sigma_z$. Linearizing about the band-touching point $k_0 = \pi/2$ ($k = \pi/2 + q$, $|q| \ll 1$) yields

$$H(q,t) = 2\tau q \sigma_x - i\gamma(t)\sigma_z, \tag{5}$$

with dispersion $E^2 = (2\tau q)^2 - \gamma^2(t)$. This admits a one-to-one correspondence with the Dirac model $\omega^2 = c^2 k^2 - (\kappa(t)c^2)^2$ under $c \leftrightarrow \tau$, $\kappa(t)c^2 \leftrightarrow \gamma(t)$, and $k - \pi/2 \leftrightarrow q$. So, the k-gap criteria are identical and a temporal-kink train $\gamma(t)$ realizes the same mass inversion used in the PTC modelling.

Using the same two-seed-pulse protocol as in the continuum, we launch identical Gaussians from opposite sides of the chain with a tunable phase difference $\Delta\phi$. Parameters are $\tau = 1.9, \gamma_0 = 0.72, T_0 = 7.5T$, and $T_r = 1.71T$ (with $T = 2\pi/\omega$). A site-resolved snapshot [Fig. 3b] contrasts three cases: $\Delta\phi = 0$ (red), $\pi/2$ (blue), and $\pi$ (black). In all cases, the lattice outside the triangular overlap region (TOR) remains nearly unchanged, indicating that phase-dependent gain or suppression is confined to the zone where the counter-propagating partners meet. The full spatiotemporal evolutions [Figs. 3c–3e] exhibit a kink-to-kink scattering cascade. The TOR intensity follows a phase-ordered hierarchy: construction (largest signal) at $\Delta\phi = 0$, near the linear-addition baseline at $\Delta\phi = \pi/2$, and strong depletion at $\Delta\phi = \pi$. These trends replicate the dynamics in the PTC picture, where each kink flips the sign of the effective mass $\gamma(t)$, converting the amplified in-gap branch into the decaying one.

**Further discussions.** The massive Dirac equation furnishes a simple, unifying description of dynamics near k-gap opening in both PTC continuum and lattice realization. In the continuous model, an imaginary mass opens a momentum gap, while a time-domain Jackiw-Rebbi kink implements a mass inversion that yields temporally growing–then–damping transient behaviors and enables phase-selective pair annihilation between two seeded pulses. In parallel, a simple time-varying non-Hermitian lattice, linearized near $k \approx \pi/2$, reduces to the equivalent Dirac Hamiltonian under the parameter map $c \leftrightarrow \tau$, $\kappa c^2 \leftrightarrow \gamma$. Correspondingly, under an idential two-seed protocal, the dynamics in the lattice recovers the similar spatiotemporal patterns with topologically protected pair annihilation.

Collectively, these results establish a one-to-one correspondence among the continuum, lattice, and Dirac perspectives: temporal modulation acts as a programmable imaginary Dirac mass, and k-gap physics is faithfully captured by the massive Dirac equation. The framework affords concrete implementation pathways and diagnostics in circuits [17,18], acoustics [19,20], and synthetic-dimension platforms [11,21,22], and it advances temporal kinks as a practical time-gating strategy for controlling wave propagation and mode growth in time-varying systems. This correspondence further provides a transferable theoretical basis for more elaborate spatiotemporal modulations from acoustics to optics.

Converging advances already realize each operation required by our protocol - momentum-gap engineering, controllable in-gap amplification, and temporal topological modes - thus supplying an experimental toolkit. Dynamically modulated transmission-line circuits enable direct measurement and modelling of a genuine k-gap [17,18]; B. Zhang et al. [23] further demonstrate the in-gap wave amplification and observe a temporal topological state (Zak-phase mid-gap mode), while resolving the growing and decaying branches. Time-varying photonic metasurfaces verify the PTC k-gap and harness its in-gap exponential amplification as an engineered platform [24], with resonant implementations broadening the accessible k-gap [25]. Temporal synthetic lattices in coupled fiber loops implement domain interfaces with distinct invariants [11,12], and PT-symmetric acoustic Floquet lattices corroborate quantized time-topological structure and support time-localized boundary modes [20]. Our suggestion is to synthesize these current capabilities into a phase-programmable topological protocol: temporal kinks applied to two seeded pulses toggles growth or decay and yields superluminal pair annihilation, thereby elevating k-gap physics to a controllable dynamical resource. Given the available ingredients across platforms, near-term experimental realization is anticipated.

**Conclusion.** In short, we proposed that a sequence of topological temporal kinks inside the k-gap can induce superluminal pair annihilation. This phenomonon arises in both a continuum photonic time-crystal and time-varying non-Hermitian lattice models, each of which maps onto a Dirac equation with a kinked imaginary mass term. The continuum–lattice correspondence provides a clear operational signature of pair annihilation, indicting temporal kinks as unified manifestations of an imaginary Dirac mass flip. We thus present temporal kinks as time-gating, making pair annihilation a controllable and experimentally accessible effect in time-varying media, and opening avenues for programmable energy exchange and spatiotemporal pattern formation across photonic, circuit, acoustic, and synthetic-dimension platforms.

**Acknowledgement**
Y. P. acknowledges the support of the NSFC (No. 2023X0201-417-03) and the fund of the ShanghaiTech University (Start-up funding).

The authors declare no competing financial interests.

Correspondence and requests for materials should be addressed to Y.P.

(yiming.pan@shanghaitech.edu.cn), and (qqcheng@usst.edu.cn)

**Methods**

**Dirac formalism from PTC.** We start by modelling a linear photonic time crystal (PTC). In a non-magnetic dielectric medium without free charges and moving currents ($\rho = 0, \boldsymbol{J} = 0$), Maxwell's equations in source-free form are

$$\nabla \times \boldsymbol{E} = -\frac{\partial \boldsymbol{B}}{\partial t}, \nabla \times \boldsymbol{B} = \mu_0 \frac{\partial \boldsymbol{D}}{\partial t}$$

with the divergence constraints $\nabla \cdot \boldsymbol{D} = 0, \nabla \cdot \boldsymbol{B} = 0$, where **E** is the electric field, **D** is the electric displacement, **B** is the magnetic field, and $\mu_0$ is the vacuum permeability. Applying the curl operator ($\nabla \times$) to the first equation and substituting the result into the second yields:

$$\frac{\partial^2 \boldsymbol{D}}{\partial t^2} = -\frac{1}{\mu_0} \nabla \times (\nabla \times \boldsymbol{E}).$$

Neglecting nonlinear contributions of media relative to the temporal modulation, the electric displacement field can be expressed as $D = \epsilon(x,t)E = \epsilon_0 \epsilon_1(x,t) E$. The linear dielectric constant, $\epsilon_1$, is modulated periodically in time, and spatially uniform

$$\epsilon_1(x,t) = \epsilon_r(1 + \delta \cos \Omega t)$$

Here, $\epsilon_r$ is the mean permittivity, $\delta \ll 1$ as a small modulation strength (typically up to ~0.3 $\epsilon_{eff}$). And $\Omega = 2\pi/T$ specify the temporal modulation frequency, with period $T$. Accordingly, a light pulse with central wavelength 800 nm requires a temporal modulation with period of $T$=2.7 fs for the setting of a PTC. Substituting into the wave function yields

$$\frac{\partial^2 D}{\partial t^2} = \frac{1}{\mu_0} \frac{\partial^2}{\partial x^2}\left(\frac{D}{\epsilon_0 \epsilon_1}\right).$$

In time-varying media, the electric displacement D is favoured over the electric field E because D remains continuous across temporal boundaries, as dictated by Gauss's law ($\nabla \cdot \boldsymbol{D} = 0$), whereas E may be discontinuous. Such temporal boundaries give rise to phenomena including time reflection and time refraction. For small modulation amplitudes $\delta \ll 1$, Eq. 1 can be simplified to:

$$\frac{1}{c^2}\frac{\partial^2 D}{\partial t^2} = (1 - \delta \cos \Omega t)\frac{\partial^2 D}{\partial x^2}.$$

Here $c = c_0/\sqrt{\epsilon_r} = c_0/n_0$ is the speed of light in the medium, with $n_0 = \sqrt{\epsilon_r}$ being the refractive index and $c_0$ the light speed in vacuum. We approximate $\tilde{\epsilon}^{-1}(t) = 1 - \delta \cos \Omega t$.

To proceed, we employ a coupled-mode framework based on the Floquet–Bloch theorem. Floquet-Bloch waves are constructed as a superposition of temporally modulated forward and backward components

$$D(x,t) = A_f(x,t)e^{ikx-i\omega t} + A_b(x,t)e^{ikx+i\omega t} + h.c.,$$

with $\omega = \Omega/2$ and $k = \Omega/2c$. Here $A_f$ and $A_b$ are the slowly varying complex amplitudes of forward- and backward-propagating modes, representing time refraction and time reflection, respectively. Substituting the ansatz into time modulated wave function and applying the slowly-varying envelope approximation ($\delta \ll 1$) yields the coupled-mode equations

$$\frac{i}{c}\frac{\partial A_f}{\partial x} + i\frac{\partial A_f}{\partial x} + \frac{\delta\Omega}{8c}A_b = 0$$

$$-\frac{i}{c}\frac{\partial A_b}{\partial t} + i\frac{\partial A_b}{\partial x} + \frac{\delta\Omega}{8c}A_f = 0$$

These equations describe the mutual coupling between the time-reflected and time-refracted modes, and allow one to focus on the regions where mode coupling is strongest to investigate gap formation. We recast the coupled-mode system into a 1+1D Dirac equation. Starting from

$$\left[\frac{1}{c}\left(i\frac{\partial}{\partial t}\right)\sigma_z + \left(i\frac{\partial}{\partial x}\right)\sigma_0 + \frac{\delta\Omega}{8c}\sigma_x\right]\psi = 0,$$

where $\psi = (A_f, A_b)^T$ and $\sigma_0$ the identity ($\sigma_{x,z}$ the Pauli matrices). Left-multiply by $\sigma_z$ (so we can use $\sigma_z\sigma_x = i\sigma_y$)

$$\frac{i}{c}\partial_t\psi + i\sigma_z\partial_x\psi + i\frac{\delta\Omega}{8c}\sigma_y\psi = 0$$

Multiplying by $c$ and moving the spatial and coupling terms to the right gives

$$i\partial_t\psi = -ic\sigma_z\partial_x\psi - i\kappa c^2\sigma_y\psi,$$

with $\kappa = \delta\Omega/8c^2$. This has the standard Dirac form.

**Reference.**

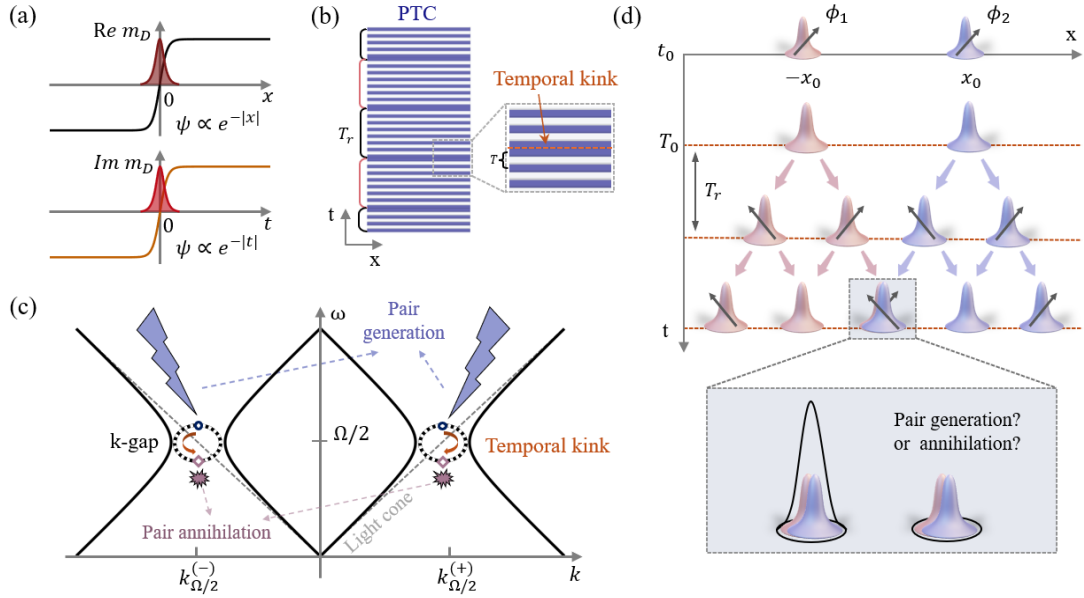

**Figure 1| Pair annihilation in the $k$-gap of a photonic time crystal driven by a temporal-kink modulation.** (a) Zero mode induced by domain wall and temporal edge state induced by temporal kink. (b) Schematic of a photonic time crystal (PTC) with temporal kinks superimposed on a time-periodic modulation of period T. (c) Floquet dispersion $\omega(k)$ of the PTC showing two symmetric $k$-gaps centered near $\Omega/2$, and pair generation and pair annihilation of the gap modes. (d) Two seeding pulses with different initial phases evaluate in PTC, with one from each overlapping at temporal kinks.

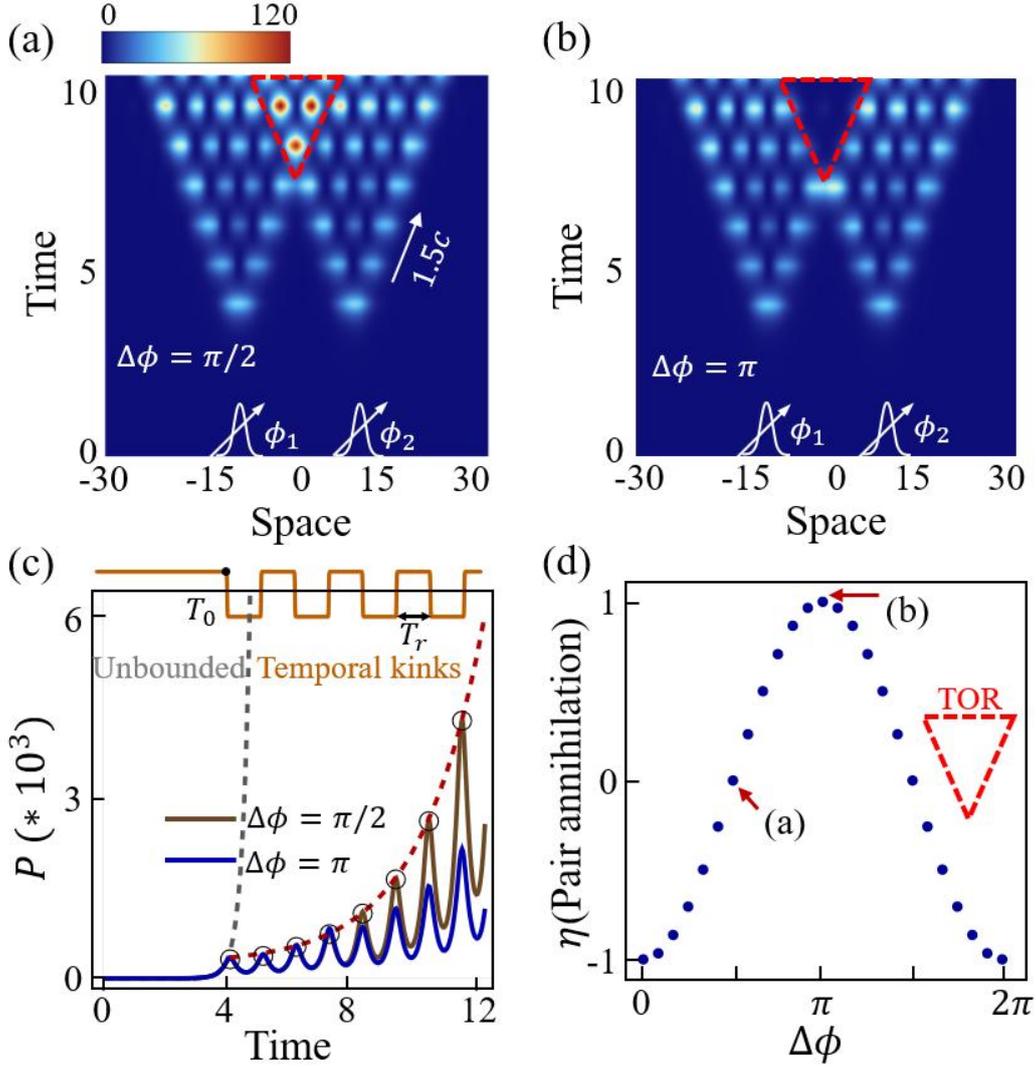

**Figure 2| Temporal kinks induce superluminal pair annihilation in photonic time crystal.** (a) Gaussian inputs launched inside the $k$-gap with initial phase difference $\Delta\phi = \pi/2$, temporal-kinks driven superluminal triangular spatiotemporal patterns, and their intensities simply superpose in the overlap region. (b) For an initial phase difference $\Delta\phi = \pi$, the two spatiotemporal patterns undergo pair annihilation in the overlap region. (c) Total energy for $\Delta\phi = \pi/2$(brown) and $\Delta\phi = \pi$(blue). Circles indicate increments at successive kinks; the red dashed line is an envelope. (d) Pair-annihilation efficiency versus phase, computed over the triangular overlap region (TOR).

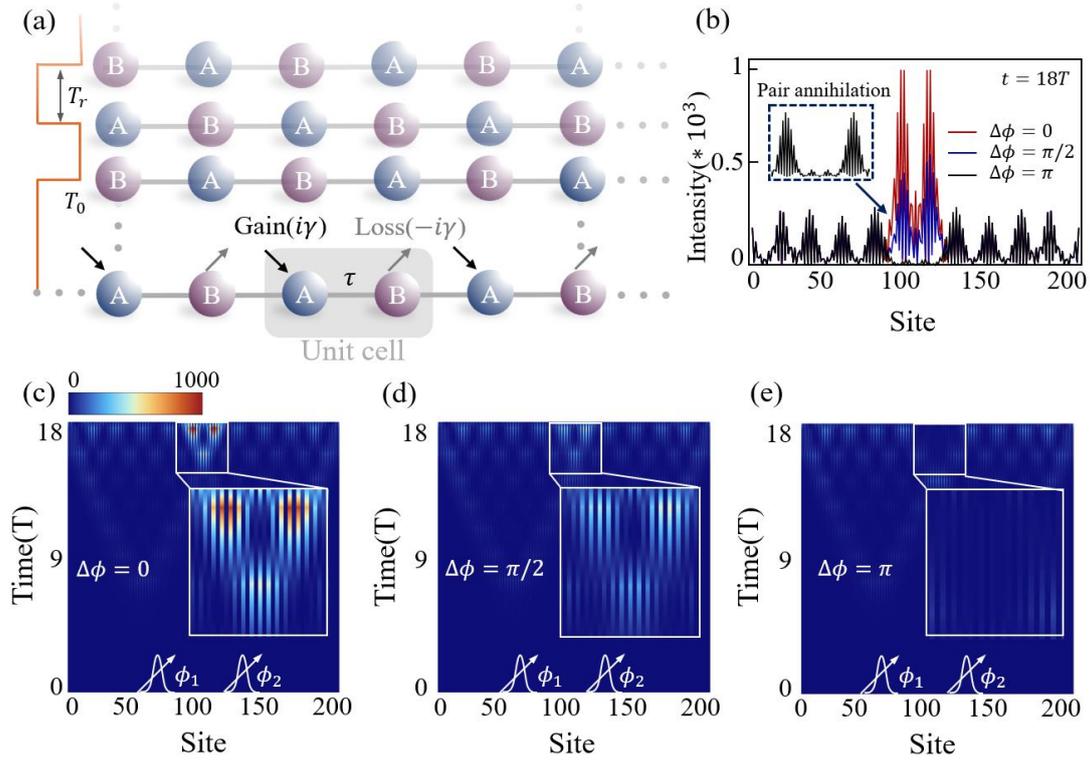

**Figure 3| A minimal lattice model with gain-loss modulation and its dynamics.** (a) Schematic of the bipartite lattice with alternating intracell hopping $\tau$ and balanced gain/loss terms ($\pm i\gamma$). (b) Site-resolved intensity at $t = 18T$ for three initial phase differences. (c, d, e) Real-time site-resolved dynamics for $\Delta\phi = 0$, $\Delta\phi = \pi/2$ and $\Delta\phi = \pi$ showing distinct phase-dependent interference, in direct correspondence with the continuous PTCs model.